# Mesoscopic Fractional Quantum in Soft Matter

W. Chen

*Institute of Applied Physics and Computational Mathematics, P.O. Box 8009, Division Box 26, Beijing 100088, P. R. China (chen_wen@iapcm.ac.cn)*

**Abstract. -** **Soft matter (e.g., biomaterials, polymers, sediments, oil, emulsions) has become an important bridge between physics and diverse disciplines. Its fundamental physical mechanism, however, is largely obscure. This study made the first attempt to connect fractional Schrodinger equation and soft matter physics under a consistent framework from empirical power scaling to phenomenological kinetics and macromechanics to mesoscopic quantum mechanics. The original contributions are the fractional quantum relationships, which show Lévy statistics and fractional Brownian motion are essentially related to momentum and energy, respectively. The fractional quantum underlies fractal mesostructures and many-body interactions of macromolecules in soft matter and is experimentally testable.**

The frequency scaling power law of fractional order appears universal in physical behaviors of soft matter (*1-5*) and is considered "anomalous" compared with those of the ideal solids and fluids. For instance, Jonscher (*4*) concluded that a fractional frequency power law was "the universal dielectric response" in soft matter. It is also well-known (*5, 6*) that acoustic wave propagation through soft matters (Fig. 1, reproduced from Ref. *5*) obeys frequency power law dissipation. The standard mathematical modeling approach using integer-order time-space derivatives can not accurately reflect fractional frequency power law, while the fractional derivatives are instead found an irreplaceable modeling approach. In particular, anomalous diffusion equation has been recognized as a master equation (*8*) to describe frequency power scaling of various physical processes (e.g., transport, relaxation) and is stated as

$$\frac{\partial^\eta s}{\partial t^\eta} + \gamma(-\Delta)^{\mu/2} s = 0, \ 0 \prec \eta \leq 1, \quad 0 \prec \mu \leq 2 \qquad (1)$$

where *s* is the physical quantity of interest, $\gamma$ the corresponding physical coefficient, $(-\Delta)^{\mu/2}$ represents the symmetric non-local positive definite fractional Laplacian, and $\eta<1$ leads to the non-local time derivative of fractional order (*7, 9*). Note that $\eta$ and $\mu$ are in general real numbers, and "fractional" in this letter is traditional misnomer in academic nomenclature. For the ideal solids and fluids (e.g., water, crystals), $\eta=1$ and $\mu=2$; while for soft matter such as polymers, colloids, emulsions, foams, living organisms, rock layers, sediments, plastics, glass, rubber, oil, soil, DNA, etc, $\mu$ ranges from 0 to 2 and $\eta$ is from 0 to 1. It is worth pointing out that equation (1) concerns with a phenomenological time-space representation but does not necessarily reflect physical mechanisms behind the scenes (*7, 8*). By using the separation of variables, namely, $s(x,t) = T(t)Q(x)$, we have

$$(-\Delta)^{\mu/2} Q - \lambda Q = 0 \qquad (2)$$

where $\lambda$ are eigenvalues of the fractional Laplacian and are also the minima of the potential energy. Applying the Fourier transform to equation (2) and $F\{(-\Delta)^{\mu/2} Q\} = |k|^\mu \hat{Q}$, we get $\lambda = |k|^\mu$, where *k* denotes the wavenumber (*9*). Now we find the discrete potential energy spectrum of fractional order $\mu$ ($\mu \neq 2$) from phenomenological master equation (1). This implies that unlike ideal solids, soft matter has the fractional energy band.

On the other hand, the fractional Schrodinger equations have recently been proposed through a replacement of time/space derivative terms in the standard Schrodinger equation by the corresponding fractional derivatives (*10-14*)

$$e^{i\pi\eta/2}\hat{h}_\eta \frac{\partial^\eta \Psi}{\partial t^\eta} = \frac{\hat{h}_\mu^2}{2m}(-\Delta)^{\mu/2}\Psi + V\Psi \quad (3)$$

where $\hat{h}_\eta$ and $\hat{h}_\mu$ are the scaled Planck constant in terms of $\eta$ and $\mu$, $\Psi$ the wave function, and *V* represents the potential energy. The fractional derivative representation underlies a dissipative process. Equation (3) is not a consequence of a basic principle of physics. Laskin's work (*13*) was inspired by Feynman's discovery that quantum trajectories are of fractal self-similarity nature. By the quantum integral over the Lévy paths in contrast to the conventional Feynman Gaussian path integral, Laskin (*13*) derived his fractional Schrodinger equation with an anomalous kinetic term $(-\Delta)^{\mu/2}$ and found the fractional energy spectrum of an artificial hydrogenlike atom, called the fractional "Bohr atom", without referring to soft matter and the fractional frequency power law. The eigenvalue equation of fractional Schrodinger equation (*13, 14*) appears essentially the same as macromechanics fractional equation (2) for the potential *V*=0, while either energy may have different physical interpretation. The potential energy *V* in equation (3) is problem-dependent (*13, 14*). Fig. 2 (reproduced from Ref. *14*) displays band structures of fractional order of fractional Schrodinger equation under a variety of potentials. Anomalous diffusion equation of power law scaling in soft matter and fractional quantum mechanics have common signature of the fractional energy band.

Statistically, anomalous diffusion equation (1) and fractional Schrodinger equation (3) are characterized by Lévy statistics (long fat tailed distribution, $\eta=1$, $\mu<2$) and the fractional Brownian motion (long-range correlation, $\eta<1$, $\mu=2$) (*15, 16*). $\mu$ is also the stability index of Lévy distribution, while $\eta$ is the index of memory strength involving correlation function of physical process (dependence on the past history of motion, non-Markovian process), and the smaller it is, the stronger memory (*17*). And we can also obtain the fractional Langevin kinetic description of power law noise and the fractional Fokker-Planck equation of the corresponding probability density function (PDF) (*16, 18*). In recent decade both "anomalous" statistic paradigms have been recognized mathematical foundation of the classical (*18, 19*) and quantum (*14, 19, 20*) statistical physics of soft matter. Given this statistic argument and the signature of fractional energy spectrum, we consider the fractional Schrodinger equation the underlying physics of soft matter under mesoscopic quantum scale, while phenomenological kinetics of the fractional Langevin and Fokker-Planck equations in the classical limit along with anomalous diffusion equation describe microscopic and macroscopic physical behaviors, respectively. The standard kinetics, macromechanics and quantum mechanics underlying Gaussian Brownian motion are the limiting cases $\eta=1$, $\mu=2$.

The kinetic energy of fractional order in Laskin's fractional Schrodinger equation was given by (also in Ref. *14*)

$$E_k = D_\mu |p|^\mu, \qquad (4)$$

where $D_\mu$ is the scaled constant with the physical dimension $\text{erg}^{1-\mu} \times \text{m}^\mu \times \text{sec}^{-\mu}$ and $p$ denotes momentum (*13, 14*). In terms of the standard quantum momentum relationship $p = hk$, we have $E_k = D_\mu h^\mu |k|^\mu$. Further comparing it with $E_k = |p|^2/2m$, we derive

$$p = \hat{h}_\mu k^{\mu/2}, \quad 0 \prec \mu \leq 2 \qquad (5)$$

where $\hat{h}_\mu = \sqrt{2mD_\mu h^\mu}$. Here we encounter a fundamental contradiction between

$E_k = |p|^2 / 2m$ and linear $p = hk$ in the presence of fractional order kinetic energy. In essence, the energy and momentum relationship (4) contradicts with $E_k = |p|^2 / 2m$ and is superficial. In contrast, $E_k = D_\mu h^\mu |k|^\mu$ is fundamental. The fractional quantum relationship (5) simply avoids the paradox inherent in (4) and saves $E_k = |p|^2 / 2m$. And (5) results only in Lévy process (fractional Laplacian) with respect to momentum and is thus considered soft matter quantum. The Fourier transform of the Lévy probability density function is $P(k) = e^{-\gamma |k|^\mu}$. In terms of the fractional quantum, the Lévy PDF of energy state is $P(E_k) = e^{-E_k / (D_\mu \hat{h}/\gamma)}$ in the classical Maxwell-Boltzmann statistics fashion. Similarly, we can derive the Bose-Einstein and the Fermi distributions corresponding to the Lévy PDF of energy state.

In the fractional Schrodinger equation (3), the fractional Laplacian corresponds to the Lévy process (*13, 14*), while the fractional time derivative representation accounts for the fractional Brownian motion (*11, 12*). Through a quantum plane wave analysis, we find the physical underpinning of the latter is a fractional Planck quantum energy relationship

$$E = \hat{h}_\eta \nu^\eta, \quad 0 \prec \eta \leq 1, \qquad (6)$$

where $\nu$ denotes the frequency. The quantum relationship (6) suggests that quantum process in soft matter only involves the fractional Brownian motion with respect to energy and also implies quantum particles of fractional order. For instance, acoustic wave propagation is actually vibration of the molecules of media, and the quantized energy of this vibration (quantum oscillator) is called phonons. In terms of $E = \hat{h}_\eta \nu^\eta$, acoustic energy is transmitted and absorbed in fractional quantum, the history-dependent fractional phonon, through soft matter, underlying anomalous vibration.

The physical principles apply on the physical size scale. Einstein's relativity theory has revealed our universe on the large scale and shown a profound link between time and space. But the theory does not hold at the microscopic subatomic level, where quantum mechanics theory prevails. In soft matter, the large amount of the elementary molecules is grouped together on mesoscopic scale, in between microscopic and macroscopic scales, and behaves like a macromolecule. Physical properties of soft matter can not arise from relativistic or quantum properties of elementary molecules (*1*), which have otherwise been extremely successful in ideal solids and fluids. Macromolecule mesostructures, average volume much larger than atomic scales, are non-local fractal and not lattice (e.g., the folding of DNA macromolecules (*21*)) and play a vital role in determining physical properties (*1*). Refs. *19* and *22* give a detailed analysis of how the fractional mechanism arises naturally from a dispersive quantum particle evolving in a chaotic environment characterized by Lévy fluctuation, analogous to mesostructures in soft matter. For instance, many-body cooperative interactions, other than the isolated polar molecules of the Debye model (independent relaxation), have been considered the physical mechanism behind soft matter dielectrics (*23*). From the fractional quantum point of view, complex many-body interactions obey Lévy statistics (similar to complex quantum system in *22*) and cause fractional band structures and determine fractional frequency power law dielectrics.

The fractional indices $\eta$ and $\mu$ through this report are actually fractal (dimension of fractional order, see Ref. *24*) and characterize affect of time-space topological complexity of macromolecules on mesoscopic quantum and macroscopic physical processes (e.g., dispersion). Both appear in fractional empirical power law scalings, kinetics, diffusion, wave, quantum, and statistics equations of soft matter, while the classical quantum and statistics ($\eta=1$, $\mu=2$) are recovered for free particles and ideal solids and fluids (atomic lattice) in which correlation decays exponentially and influence of the past is very weak. In a descriptive level, Lévy statistics and fractional Brownian motion produce common characteristic process, such as the power-law growth of the second moment and long-range correlation (algebraic decay), but are fundamentally different (*21*), respectively related to superdiffusion ($\eta=1$, $\mu<2$) and subdiffusion ($\eta<1$, $\mu=2$)(*18, 19*). Both "anomalous" statistics have in recent years been observed in a wide variety of quantum systems (e.g., quantum excitation, tunneling, laser cooling (*25*)) and experiments (*20, 26 , 27*), involving amorphous semiconductor, polymer, porous media, quasicrystals, and fractal lattices, see, for example, Ref. *11* and references therein.

In summary, this study explored the links between fractional power law scaling, anomalous diffusion, fractional kinetics, Lévy statistics, fractional Brownian motion, fractal topology, fractional derivatives, and fractional quantum in soft matter and incorporated the previous results in scattered reports under a consistent theoretical framework. The original contributions of this work are the new mesoscopic fractional quantum relationships (5) and (6), which can derive dissipative fractional Schrodinger equation (3). This study also made the first attempt to connect fractional Schrodinger equations and soft matter physics through statistic arguments and the common signature of macromechanics and mesoscopic quantum. Although the connection is presented in somewhat heuristic way and need further be solidified in the future research, the fractional quantum of soft matter is physically sound and mathematically consistent and underlies mesostructures and many-body long-range interactions of macromolecules such as entanglements, branching and cross-linking, essentially responsible for universal power law scaling. The fractional energy band structures and quantum relationships can be tested in experiments through measurements of excitation, vibration, and absorption spectra of such soft matters as polymers, oil, human tissues, DNA macromolecules, or emulsions, in which parameters $\eta$ and $\mu$ can be determined.

The classical physics is mostly linear in its principle which describes and explains successfully behaviors of microscopic atomic lattices and free particles, while the fractional physics has the nonlinear physics principle caused by fractal mesostructures. For instance, the fractional Laplacian modeling underlies the concept of the fractional Riesz potential stemming from topological complexity (*6*, *9*). It is stressed that the non-linearization of physics equations of integer-order derivatives alone can not accurately describe fractional power law scaling and various "anomalous" quantum phenomena (versus fractionalization). In literature Lévy statistics and fractional Brownian motion arise mostly in physics at a descriptive level. In contrast, either statistics is now a direct consequence of the basic fractional quantum relationships, which essentially represent history-dependent and non-local fractal topological properties (e.g., non-isolated quantum systems in Ref. *22*) of collective behaviors. As of the future quantum physics, Sir Atiyah (*28*) once pointed out "Perhaps we need to know the past in order to predict the future, perhaps the universe has memory, perhaps laws of physics are governed by integro-differential equations involving integration over the past", of which my interpretation is fractional time derivative equations underlying long-range correlation inherent in the fractional Brownian motion. In particular, he (*28*) stresses the role of topology in quantum theory, whose implication in this work is the influence of fractal mesostructures on quantum mechanics. Westerlund (*29*) describes an abundance of natural phenomena, from weather prediction to finance to geophysics, to just name a few, in which the past can not simply be truncated in the prediction of the future. This research also shows that the fractional derivative models can lead us to the deep portions of the physical mechanism. As Baglegy and Torvik (*30*) put it, the fractional calculus equation representations "should be viewed as something more than an arbitrary construction which happens to be convenient for the description of experimental data".

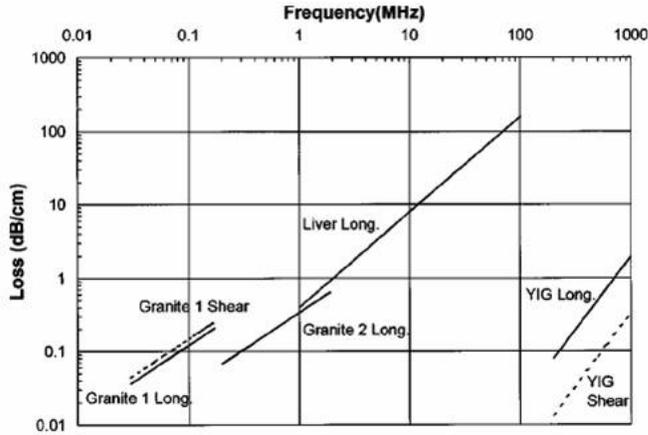

Fig. 1. Data for shear and longitudinal wave loss which show power-law dependence over four decades of frequency (taken from *5*). Acoustic attenuation typically exhibits an empirical frequency dependency characterized by a power law function of frequency, $\alpha_0 |\omega|^\mu$, where ω denotes frequency, $\alpha_0$ and $\mu \in (0,2]$ are media-specific parameters. This figure shows log-log plots of absorption versus frequency in some materials, where "shear" and "long" means respectively shear and longitudinal waves. YIG is the abbreviation of yttrium indium garnet, and granites 1 and 2 denote the two types of granite, respectively. The unit decibel (dB) is based on powers of 10 (decade) to provide a relative measure of the sound intensity. The slope of the straight line is the exponent of frequency power law of dissipation. For example, $\mu$=1.3 for 1–100 MHz in longitudinal wave loss of bovine liver. YIG as a single crystalline material has $\mu$=2 for both longitudinal and shear absorptions at very high frequencies. Clearly, YIG is an ideal solid (atomic lattice) rather than soft matter (fractal macromolecules). The longitudinal wave dissipation of granite 1 follows a linear dependence ($\mu$=1 corresponding to the fractional Laplacian of 1/2 order (*7*)) on frequency from 140 Hz to 2.2 MHz.

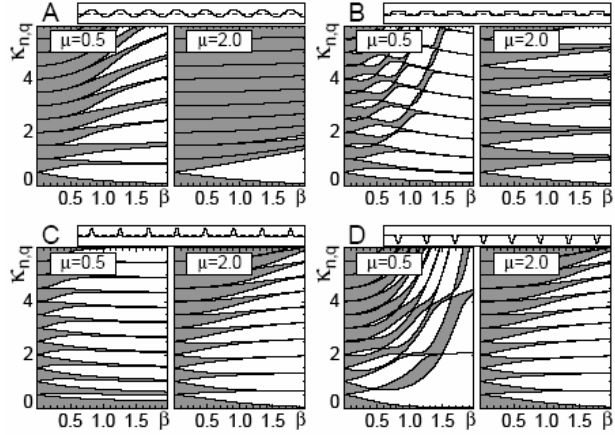

Fig. 2. Band structures of fractional order of folded polymers (taken from *14*). $\kappa$ (equivalent to *p* in this study) represents momentum, $\beta$ denotes intensive inverse temperature, *q* is the continuous Bloch phase and *n* the discrete band index, and *A*, *B*, *C*, *D* in four panel stand for four different potentials (cosine potential, square potential, localized high potential barriers and wells). The value of $\mu$ is determined by the folding topological properties of the polymer. Ref. *14* chose momentum instead of energy to express band structures but both quantities were incorrectly related by $\kappa = E^{1/\mu}$ (see the fractional quantum relationship (4) in this report). Despite this error, the figures nevertheless manifest sharp variances schematically of band structures between the normal Laplacian ($\mu$=2) and the fractional Laplacian ($\mu$=1/2) representations in Schrodinger equation.